\documentclass[10pt,aps,prb,article,twocolumn]{revtex4}
\usepackage{amsmath}    
\usepackage{amsfonts} 
\usepackage{amssymb}
\usepackage{latexsym}
\usepackage{graphics}
\usepackage{epsfig}
\usepackage{graphicx}
\usepackage[compact]{titlesec}
\titlespacing{\section}{0pt}{2ex}{2ex}

\begin{document}
\title{Optical reflectivity and magnetoelectric
effects on
resonant plasmon modes in composite
metal-multiferroic
systems}
\author{H. Vivas C.}
\affiliation{Departamento de F\'{i}sica,
Universidad
Nacional de Colombia, Sede Manizales, A.A. 127,
Col.}
 \email{hvivasc@unal.edu.co} 
\author{C. Vargas-Hern\'{a}ndez}
\affiliation{Grupo de las Propiedades \'{O}pticas
de los
Materiales (POM)\\ Departamento de F\'{i}sica,
Universidad
Nacional de Colombia, Sede Manizales, A.A.
127, Col.}
\date{\today}
\begin{abstract}
The r\^{o}le of the magnetoelectric effect upon
optical
reflectivity is studied by adapting an
electrodynamic-based model for a system composed
by a 2D
metallic film in contact with an extended
multiferroic
material exhibiting weak ferromagnetism. The
well-known
\emph{Nakayama's} boundary condition is
reformulated by
taking into account the magnetoelectric coupling
as well
as an externally applied magnetic field
$\mathbf{B}$ in an
arbitrary direction. It is found that the reflectance shows strong fluctuations for incident radiation close to the characteristic antiferromagnetic resonance frequency associated with the multiferroic material in the THz regime. These results were verified for a 10 nm metallic foil by using a finite element method (FEM) and the Rouard's approach, for a wide range of wavelengths (0.1 - 5 mm), showing good agreement with respect to Nakayama's outcome, for the particular material BaMnF$_4$. 
\\
\\
\emph{Keywords:} Multiferroics, Magnetoelectric
effect, Surface Plasmon, Reflectance.
\newline
\leftline{DOI:\hspace{2cm}PACS numbers: 73.20.-e,75.82.+t, 78.20.Bh, 78.66.Bz}
\end{abstract}
\maketitle
\section{INTRODUCTION}
Magnetoelectric (ME) effects in multiferroic (MF)
or ferromagnetic (metallic) films have brought
remarkable interest since promising technological
applications in spintronics and ultrafast electric field control
on magnetic data storage are seen as
imminent\cite{Duan},\cite{Gerhard}.
Characterization of
the relative strength for the ME coupling can be obtained
by implementing terahertz spectroscopy in rare earth
manganites of the type RMnO$_3$ (R=Tb, Gd, Dy,
Eu:Y)
\cite{Pimenov},\cite{Talbayev},\cite{Nemec},\cite{Pimenov2}
demonstrating that the generated electromagnons
(mixed
spin-waves and photon states) represent, among
others, the
signature of the ME effect for an approximate
range of
frequencies between 10 cm$^{-1}$ and 40 cm$^{-1}$
at
temperatures where antiferromagnetic resonance
modes
(AFMR) coexist, or more recently, the key
mechanism for
controllable magnetochromism in
Ba$_{2}$Mg$_{2}$Fe$_{12}$O$_{22}$ hexaferrites
\cite{Kida}.
The magnetoelectric effect emerges when a
magnetic field
$\mathbf{H}$ can induce a polarization vector
$\mathbf{P}$ at zero applied electric
field
($\mathbf{E}=0$). Likewise, the magnetization of
the
substance $\mathbf{M}$ can be generated for an
electric
field $\mathbf{E}$ with $\mathbf{H}=0$. The
minimal
coupling for describing the thermodynamic
potential
associated with this effect is given by
$\Phi=-\alpha_{ij}E_{i}H_{j}$, where
$\alpha_{ij}$ is an
unsymmetrical magnetoelectric tensor, whose
components
depend on the magnetic symmetry
class\cite{Landau}.
The primary origin for the ME coupling is
commonly
associated with the Dzyaloshinskii-Moriya
relativistic
exchange-interaction
\cite{Nagaosa},\cite{Katsura0} which
is appropriate for the description of asymmetric
spin
wave dispersion on double layer
Fe-films\cite{Zakeri} as
well as for those materials where weak
ferromagnetism
emerges, namely the ilmenite FeTiO$_3$,
TbMnO$_3$,
Eu$_{1-x}$Y${_x}$MnO$_3$ ($0<x\lesssim 0.3$ at
$T<40$
K\cite{Mukhin}) or the widely studied
pyroelectric
ferromagnet BaMnF$_4$ \cite{Scott}. Weak
ferromagnetism on
this compound is generated by canting effects
between
antiferromagnetic sub-lattices, leading to a
spontaneous
polarization $\mathbf{P}$ perpendicular to the
resulting
magnetization $\mathbf{M}$\cite{Gun}.
Considerations in
the symmetry change of the static polarization
and
magnetization fields have brought interesting
unconventional optical phenomena labeled as
non-reciprocal
dichroism associated with the sign reversal of
$\textbf{P}\times\mathbf{M}$, recently reported
in the
perovskite Eu$_{0.55}$Y$_{0.45}$MnO$_{3}$, with
magnetoelectric activity for photon energies
around 0.8
meV (sub THz regime) in the cycloidal phase at 4
K\cite{Taka}. Intense activity in the last decade
has also
been dedicated to achieve possible optical and
photonic
band gap control via Surface Plasmon (SP)
propagation in
periodic arrays\cite{Barnes}, since modern
lithographic
techniques allow to design functional objects
with almost
any desirable geometrical pattern at a sub-wavelength
scale\cite{grooves}. Plasmon localization and its
coupling
with incident light depend on the dielectric
properties of
the metal in conjunction of its surrounding
environment,
enlightening an alternative route for engineering
highly
efficient SP photonic devices via externally
applied
fields, rare earth doping or electron charge
transference
from the modified metal\cite{Freund}. In this
communication, we study an electrodynamic-based
model for
estimating the optical response generated by the
contact
between a material exhibiting weak ferromagnetism
in
contact with a 2D metallic film. It is found that
a
specific strength of the ME interaction might
couple with
localized charge-sheet modes for electron carrier densities about $10^{14}-10^{15}$ cm$^{-2}$ and
incident
frequencies around 18 cm$^{-1}$, leading to a
change in
the reflectance from the metallic film. Applied
magnetic
field effects on relative reflective are discussed in
section III.
\section{MODEL}
\begin{figure}[!ht]
\centering
\includegraphics[width=8cm,
scale=1]{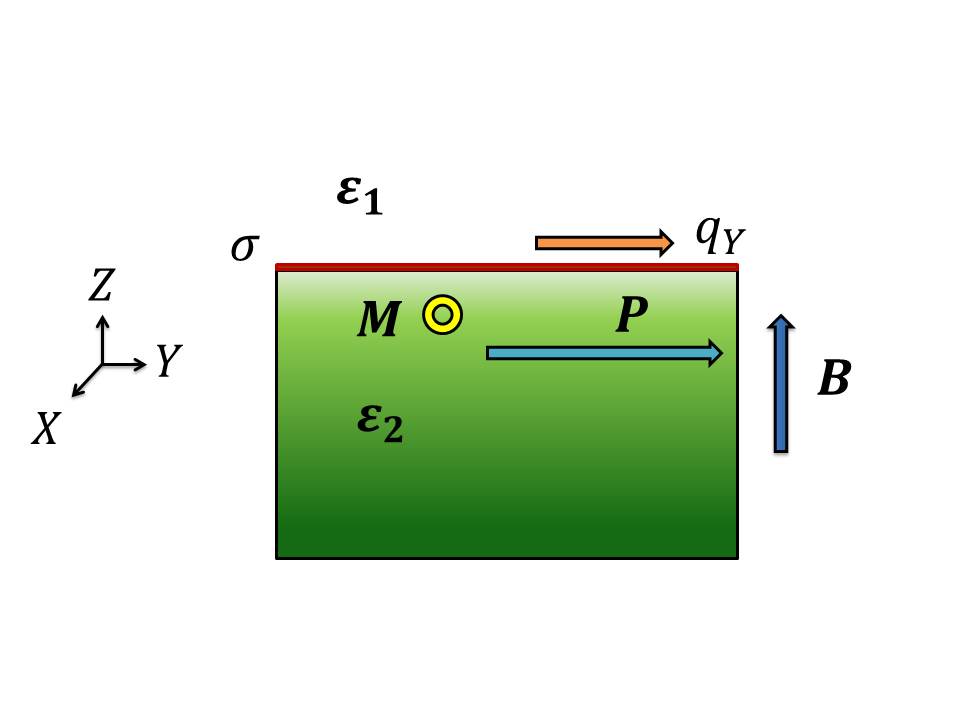}\label{fig1}
\vspace{-0.75cm}
\caption{Conducting Charge-Sheet in contact with
a
multiferroic surface. The polarization vector
$\mathbf{P}$
and the wavevector of coupled excitations $q_Y$
are also
depicted in the diagram. \emph{Weak
ferromagnetic}
magnetization vector $\mathbf{M}$ is produced by
interacting antiferromagnetic sublattices with
relative
canting angle $\theta_{C}$.}\label{r0}
\end{figure}

Localized charge-sheet modes in a 2D conducting
medium in
the framework of Drude approximation is obtained
from the
Nakayama result
\cite{Nakayama},\cite{Cottam},\cite{Pitarke}:
\begin{equation}\label{Nakay}
\frac{\varepsilon_{1}}{\kappa_{1}}+\frac{\varepsilon_{2}}{\kappa_{2}}=-\frac{ic^{2}\sigma}{\omega}=\frac{\Omega_{S}c}{\omega^{2}},
\end{equation}
where $\kappa_{j}$ corresponds to the quasiwavevector in the
$Z-$
direction, $\Omega_{S}$ is defined as $\nu
e^{2}/\varepsilon_{0}mc$ and $\nu$ denotes the electron
density concentration in a two dimensional space. $\kappa_j$ is related with the wavevector along $Y$-direction through $\kappa_{j}=\left(q_{Y}^{2}-\varepsilon_{j}\omega^{2}/c^{2}\right)^{1/2}$,
($j=1,2$). The term
$\varepsilon_{j}$ represents the relative
dielectric function value for $j$-th medium, with $\varepsilon_{1}=1$ for vacuum. In the range
of
wavelengths behind the far infrared radiation
($<1$ mm),
the dielectric function approaches to the well
recognized Lyddane-Sachs-Teller (LST) relationship:
$\varepsilon_{2}\approx
\left(1+\chi_{\infty}\right)\left(\omega_{L}/\omega_{T}\right)^{2}$,
where $\chi_{\infty}$ corresponds to the
dielectric
permittivity of the medium $j=2$ and
$\omega_{L,\left(T\right)}$ represents the
longitudinal
(transverse)-optical phonon frequency. For
numerical
purposes, we have set
$\left(\omega_{L}/\omega_{T}\right)^2\approx
1.07$, which
coincides with the relationship for the $b$-axis
normal
phonon modes in BaMnF$_4$. The permittivity
$\chi_{\infty}$ is a functional depending on
mechanical
strain deformations and polarization field
depletion in
the proximities between the multiferroic slab and
metal
film\cite{Alpay}, and is taken as constant for
zero
applied (electric) field and fixed temperature.
Formula
(\ref{Nakay}) is derived by solving the complete
set of
Maxwell equations with normal (TM wave) incidence
for
$Z>0$, and boundary conditions on the plane $Z=0$ with the
\emph{ansatz} for propagating fields
$\mathbf{E},\mathbf{H}\sim e^{-i\left(q_{Y}Y-\omega
t\right)}$ in the
region $Z=0$. Magnetoelectric effects are taken
into
consideration throughout the transverse
susceptibility
$\chi^{me}$ and the electric displacement vector
$\mathbf{D}$ is written into the constitutive
equation
like
$\mathbf{D}=\varepsilon_{2}\mathbf{E}+4\pi\chi^{me}\mathbf{H}$.
After inserting the additional term
$4\pi\left[\chi^{me}\mathbf{H}\right]$, the expression
(\ref{Nakay})
shall be modified under $\kappa_{2}\rightarrow
\kappa_{2}+
4\pi i\omega\chi^{me}/c$. In the plane $Z=0$, and
in
agreement with the geometrical configuration
shown in Figure (\ref{r0}), the non-zero surface current
density
component is defined as $J_{Y}=\sigma E_{Y}$,
where
$\sigma$ corresponds to the $\sigma_{YY}$-element
of the
generalized conductivity tensor\cite{Solyom}, and
$E_{Y}$
is the electrical field propagating on the $Y$
direction.
The generic expression for the transverse
susceptibility
$\chi^{me}$ is obtained from first principles by minimizing the free-energy density functional $\Phi$, which contains the two sublattice magnetizations, the polarization as well as external fields.  
\cite{Tilley},\cite{Stamps}. It can be summarized as:
$4\pi i\omega\chi^{me}/c=2\pi i c
g\omega\left[\left(\omega_{p}^{2}-\omega^{2}\right)^{-1}-\left(\omega_{m}^{2}-\omega^{2}\right)^{-1}\right]$,
where
$g\equiv g\left(\theta_{C},
\mathbf{M},\mathbf{P},\omega_{m},\omega_{p}\right)$
is a
coupling parameter which is an involved function
of the
canting angle between two adjacent
(antiferromagnetic)
sublattices, the spontaneous magnetization
$\mathbf{M}$
and the polarization vector $\mathbf{P}$, as well
as the
parameters $\omega_{m\left(p\right)}$. Factor $g$ is defined in terms of the
characteristic magnetoelectric frequency $S_{me}$
as $g=8\pi^{2}S_{me}^{2}/c^{2}$, given in units
of mm$^{-2}$ all throughout this
paper\cite{Rivera}, in concordance with the
spectral
weight intrinsically associated with the fitting
procedure
for the transmittance spectra via Lorentzian
model in
various multiferroic species, namely RMn$_2$O$_5$(R:Y,Tb), TbMnO$_{3}$ or LuMnO$_3$\cite{Sushkov},
and its
dependence with the externally applied magnetic
field has
been neglected for small canting angles (See for
instance
Eqs. (38) and (47) in reference [14]. Two
main poles
are clearly identified for $\chi^{me}$: the
optical
antiferromagnetic resonance mode (AFMR)
$\omega_m$ and the
soft-phonon along $\mathbf{M}$ with resonance
frequency
$\omega_{p}$, with
$\omega_{p}>\omega_{m}$. Classical plasmon
excitations in
low 2D carrier electron density are
experimentally
detected and theoretically estimated for
wavevectors
$q\lesssim 1.4$ cm$^{-1}$ and energies
$\hbar\omega\lesssim
0.5$ meV\cite{Sarma},
\cite{Chulkov0},\cite{Kanjouri},
therefore the condition
$q_{Y}^{2}>>\varepsilon_{j}\omega^{2}/c^{2}$
remains valid
in the range of interest, and the dispersion
relationship
for the coupled magnetoelectric plasma mode is
obtained by
solving the modified equation (\ref{Nakay}):
\begin{equation}\label{cuasiwv}
q_{Y}^{\pm}=\frac{1}{2}\left[Q\pm\sqrt{Q^{2}-\gamma_{2}\left(\frac{\omega}{2\pi
c}\right)^{2}\left(Q-\gamma_{1}\left(\frac{\omega}{2\pi
c}\right)^{2}\right)}\right],
\end{equation} 
with $Q=4\pi
i\omega\chi^{me}/c+\gamma_{1}\left(\omega/2\pi
c\right)^2$, $\gamma_{1}=4\pi^{2}
c\left(\varepsilon_{1}+\varepsilon_{2}\right)/\Omega_{S}$
and
$\gamma_{2}=16\pi^{2}c\varepsilon_{1}/\Omega_{S}$.
For
$\chi^{me}=0$, i.e., no magnetoelectric effects
taken
under consideration, we reproduce the expression
for the
localized plasmon mode \cite{Nakayama}:
\begin{equation}\label{qy}
\omega=\sqrt{\frac{4\pi^{2}c^{2}q_{Y}}{\gamma_{1}}},
\end{equation}
where ($+$) sign in equation (\ref{cuasiwv}) has
been
selected. Complex index of refraction
$\check{n}\left(\omega\right)$ is directly
estimated from
the wavenumber \cite{fowles}
$q_{Y}:$
$\check{n}\left(\omega\right)=cq_{Y}\left(\omega\right)/\omega$.
The lowest-order reflectance coefficient
$R\left(\omega\right)$ for normal incidence is
defined as
$R\left(\omega\right)=\mid
\check{n}\left(\omega\right)-1\mid^{2}/\mid\check{n}\left(\omega\right)+1\mid^{2}$
and its numerical profile discussed on the next
section.
$\check{n}\left(\omega\right)$ can be considered
as the \emph{effective} index of refraction for
the composite 2D metallic foil in contact with a
multiferroic (ferroelectric) system under
normal incidence of a electromagnetic wave
oscillating in
the THz regime. Applied magnetic field $\textbf{B}$ along $Z$-direction enters into
the formalism by taking symmetry considerations upon
the dependence of the electrical conductivity as a
function of $\textbf{B}$ under the transformation
$\sigma\rightarrow\sigma\left(B\right)$, with
$\sigma\left(B\right)=i\Omega_{S}c^{-1}\omega\left(\omega^{2}-\omega_{B}^{2}\right)^{-1}$.
Expression (\ref{cuasiwv}) may be reconstructed as:
$q_{Y}^{\pm}=$
\begin{equation}\label{cuasiwv2}
\frac{1}{2}\left[Q^{\prime}\pm\sqrt{Q^{\prime
2}-\gamma_{2}\frac{\left(\omega^2-\omega^{2}_{B}\right)}{\left(2\pi
c\right)^{2}}\left(Q^{\prime}-\gamma_{1}\frac{\left(\omega^2-\omega^{2}_{B}\right)}{\left(2\pi
c\right)^{2}}\right)}\right],
\end{equation} 
with
$Q^{\prime}=Q-\gamma_{1}\omega_{B}^{2}/\left(2\pi
c\right)^{2}$. The classical localized
magnetoplasmon mode
(\ref{qy}) is rewritten for $g=0$ and under $B$ like\cite{Eriksson}:
\begin{equation}\label{qmy}
\omega=\sqrt{\omega_{B}^{2}+\frac{4\pi^{2}c^{2}q_{Y}}{\gamma_{1}}},
\end{equation}
in similarity with the result (\ref{qy}). In this
particular
case the antireflective condition
($\check{n}=1$) depends
on the external magnetic field intensity 
$\lambda_{c}^{-1}=\pi/\gamma_{1}+\sqrt{\left(\pi/\gamma_{1}\right)^{2}+\left(\omega_{B}/2\pi
c\right)^2}$, which leads to a quadratic
correlation $\lambda_{c}^{-1}\propto B^{2}$ for
$\gamma_{1}\omega_{B}/2\pi^{2}c<<1$. For an arbitrary orientation of
$\mathbf{B}$, equation (\ref{Nakay}) shall be
modified on
its right side accordingly
$\Omega_{S}c\omega^{-2}\rightarrow
\Omega_{S}c\left(\omega^{2}-\omega_{B}^{2}\right)^{-1}F\left(n_{X},n_{Y},n_{Z}\right)$,
where $F\left(\cdot\right)$ is a function of the
directors
$n_{X,Y,Z}$\cite{magf}. Optical reflectivity response for
this structure might also be verified by adapting
the Rouard  method\cite{Lecaruyer},\cite{Heavens}:
\begin{equation}\label{Rou}
R_{Rouard}=\frac{r_{1-2}+r_{2-3}e^{-2i\delta}}{1+r_{1-2}r_{2-3}e^{-2i\delta}},
\end{equation}
where $r_{i-j}$ corresponds to the internal
reflectivity
between media labeled $i$ ($j$) and $\delta$ is
the
phase difference on the second medium with
thickness $\ell$, defined as
$\delta=2\pi\check{n}_{2}\ell\lambda^{-1}$. The
index of
refraction $\check{n}_{2}$ is a function of the
components
for the conductivity tensor $\left[\sigma\right]$ depending on the incoming electromagnetic field polarization. In this particular case, it is calculated as:
\begin{equation}
\check{n}_{2}=\sqrt{1+\left(i\sigma_{YY}/\omega\varepsilon_{0}\right)},
 \end{equation}
while $\sigma_{YY}$ is explicitly given by 
\begin{equation*}
i\sigma_{YY}/\omega\varepsilon_{0}=-\omega^{2}_{P}\left(\omega^{2}-\omega^{2}_{B}\right)^{-1}\left(1-\omega^{2}_{B}n_{Y}^{2}/\omega^{2}\right),
\end{equation*}
where $\omega_P$ represents the electronic plasma
frequency for the \emph{bulk} system, which is related to $\Omega_S$ through
$\omega_{P}^{2}=cN\Omega_{S}/\nu$ where $N$ being the
volumetric electron density concentration. Reference values for plasma frequencies were taken as $\omega_{P}=2.15\times 10^{15}$ Hz and $\Omega_{S}=2.12\times 10^{12}$ Hz for gold (Au) in the framework of the Drude model fitting\cite{ws}.
Factors $r_{i-j}$ in formula (\ref{Rou}) are
given explicitly by
$r_{1-2}=\left(1-\check{n}_{2}\right)/\left(1+\check{n}_{2}\right)$,
and
$r_{2-3}=\left(\check{n}_{2}-\check{n}_{3}\right)/\left(\check{n}_{2}+\check{n}_{3}\right)$,
with
$\check{n}_{3}=\sqrt{1+\left(4\pi\chi^{me}\right)^{2}}$.
Indeces of refraction are directly obtained by
reconstructing the set of Maxwell equations on
each material media. In the general case, taking
into account the ME effect in the formalism by
inserting the tensor
$\left[\mathbf{\chi}\right]$, the propagating
electric field $\mathbf{E}$ must satisfy:
\begin{eqnarray}\label{Max}
\left(\nabla\times\nabla\times\mathbf{E}\right)_{M-MF}=i\omega\mu_{0}\left(\left[\mathbf{\sigma}\right]\mathbf{E}\right)_{M-MF}\\ \nonumber +\omega^{2}\mu_{0}\mathbf{D}_{M-MF}+4\pi\omega\nabla\times\left(\left[\chi\right]\mathbf{E}\right)_{M-MF},
\end{eqnarray}  
where $\left[\mathbf{\sigma}\right]$ is the
conductivity tensor, and $\mathbf{D}$ previously
defined as the electric displacement vector, and subscript $M-MF$ indicates the region where fields propagation are evaluated, namely the metal (M) or multiferroic (MF) slab.  
\section{RESULTS AND DISCUSSION}
\begin{figure}[!ht]
\centering
\includegraphics[width=9cm,
scale=1]{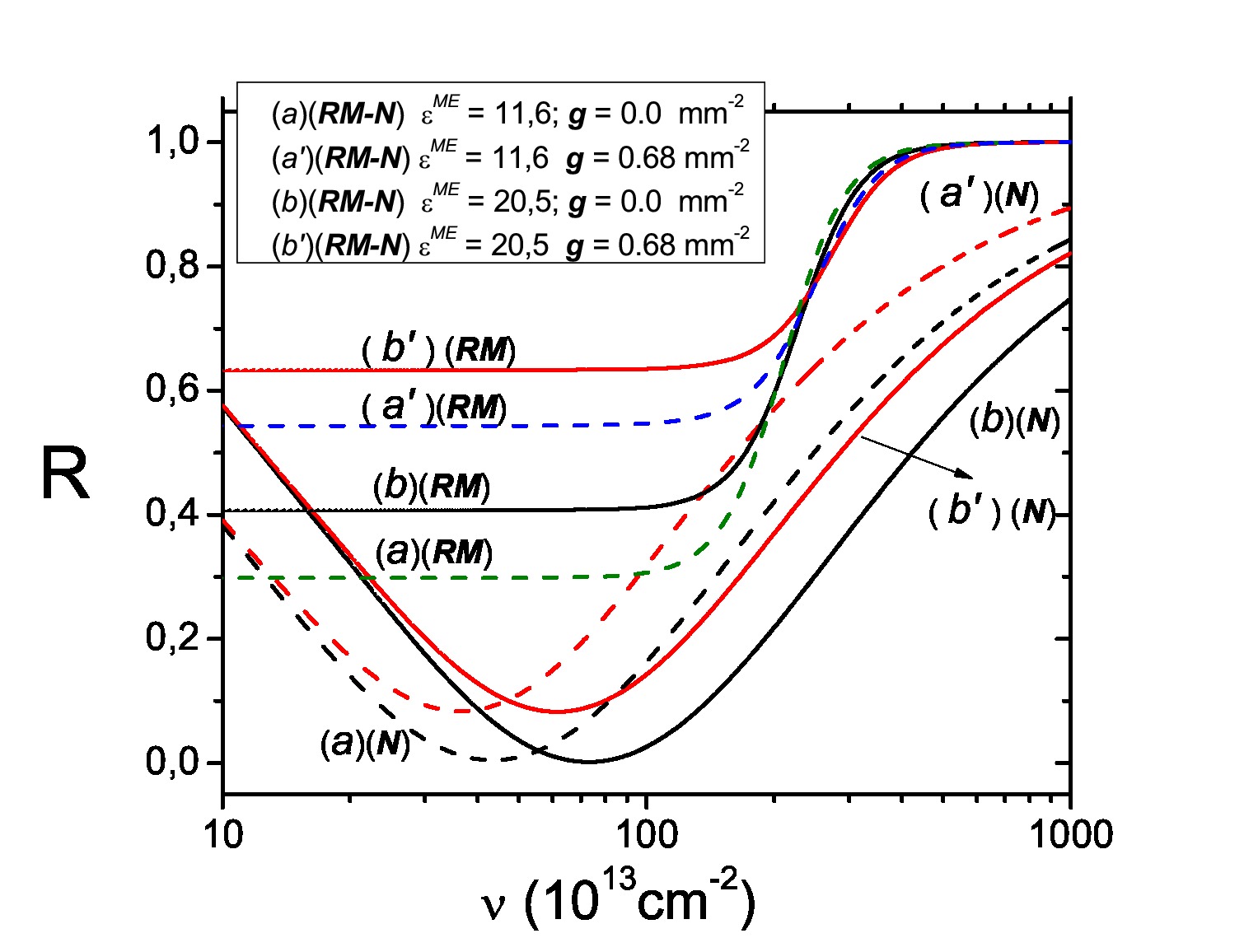}
\caption{Zero field reflectance response as a
function of
electron carrier density $\nu$, comparing the Rouard's method and Eq. (\ref{Nakay}) for the dielectric constants $\varepsilon^{ME}=11.6\varepsilon_{0}$ (curves (a) and (a$^\prime$)) and $\varepsilon^{ME}=20.5\varepsilon_{0}$ (curves (b) and (b$^\prime$)), with $\omega\approx \omega_{m}$ in all cases.}\label{r1}
\end{figure}

\begin{figure}[!ht]
\centering
\includegraphics[width=9cm,
scale=1]{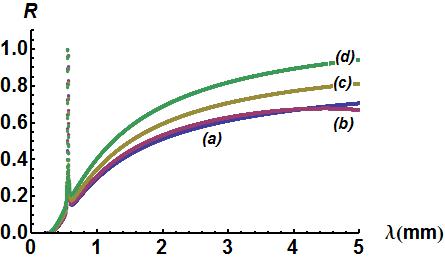}
\caption{\emph{In plane} magnetic field effects on the reflectance spectrum. $B=1.5$ T, $\theta=\pi/2$ (a) $\phi=0$, (b) $ \phi=\pi/4$, (c) $\phi=\pi/2$, (d) $\phi=\pi$, $\nu=147.42$ cm $^{-2}$, $g=0.6878$.}\label{bparallel0}
\end{figure}

\begin{figure}[!ht]
\centering
\includegraphics[width=9cm,
scale=1]{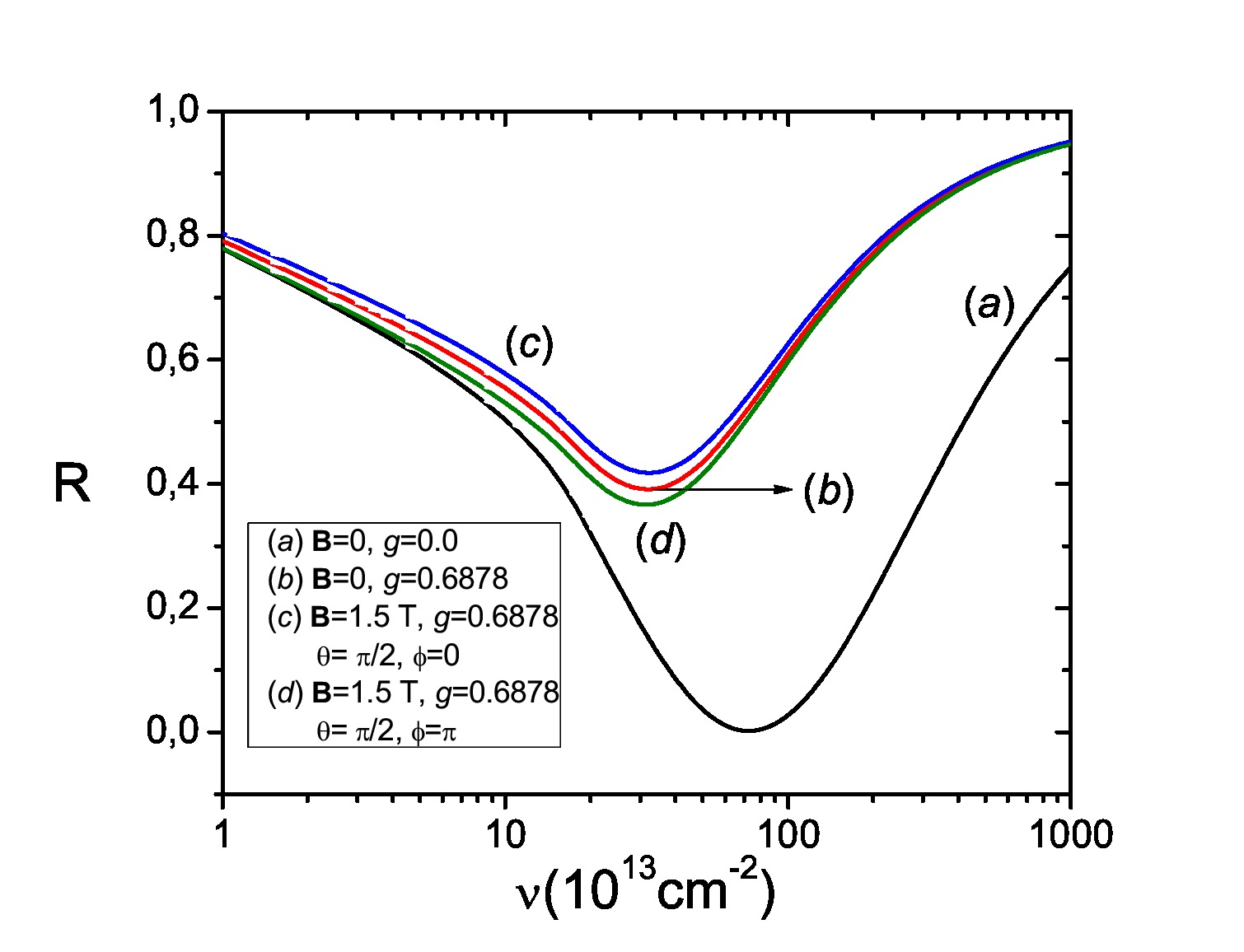}
\caption{Reflectance response as a function of electron carrier density for \emph{in-plane}
applied field close to AFMR frequency at
$2\pi c/\omega=0.53$ mm, for coupled ($g=0.6878$) and uncoupled ($g=0$) ME interaction.}\label{bparallel}
\end{figure}

\begin{figure}
\includegraphics[width=8.5 cm,
scale=1]{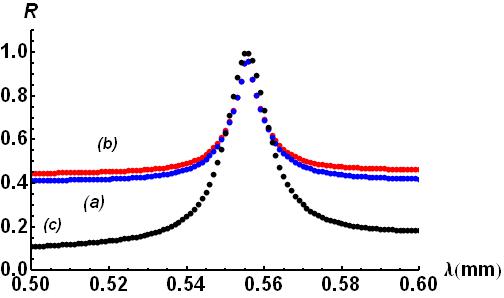}
\caption{Calculated reflectance $R$ at zero field
as
function of the incident wavelength by using
three
different techniques: (a) FEM (b) Summation
(Rouard's)
method and (c) expression (\ref{Nakay}), with an
electronic density $\nu=147.42\times 10^{13}$
cm$^{-2}$,
which corresponds to a film of $\ell\sim 10$nm
thickness.
}\label{comp}
\end{figure}

\begin{figure}
\includegraphics[width=7 cm,
scale=1]{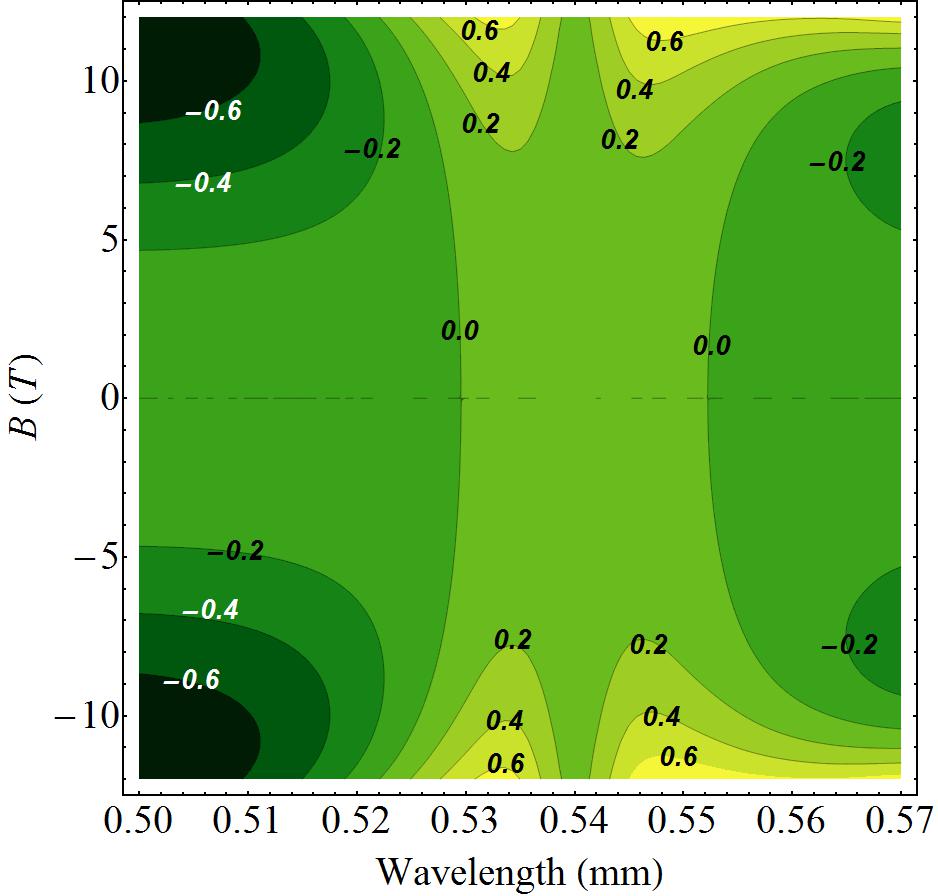}
\caption{Isoreflective lines of $\Delta R/R$
under applied
magnetic fields parallel to $Z$-axis with $\nu=0.52\times 10^{15}$ cm$^{-2}$,
$g=0.6878$ mm$^{-2}$ and $2\pi c/\omega_{m}=0.54$
mm.}\label{control}
\end{figure}

\begin{figure}[!ht]
\centering
\includegraphics[width=8.5 cm,
scale=1]{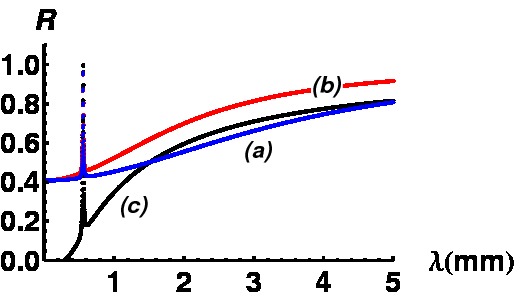}
\caption{Reflectance spectrum calculated by using (a) Finite Element Method (FEM) (b) Rouard Method (RM) and (c) Nakayama equation (N), for an applied magnetic field on $Y$- direction and $1.5$ T intense. Thickness of $\ell~\sim 10$ nm was taken in cases (a) and (b) corresponding to carrier densities $\nu=147.42\times 10^{13}$ cm$^{-2}$.  Methods (a) and (b) have good agreement at $\lambda_m$ though response (c) tends to match (a) and (b) for wavelengths $\leq 0.1$ mm.}\label{3cc}
\end{figure}

Figure (\ref{r1}) exhibits the zero field reflectance
response as a
function of the 2D electronic carrier
concentration $\nu$,
for different wavelengths and the
magneto-electric coupling parameter $g$ fixed at
$0.6878$ mm$^{-2}$, the dielectric permittivity values have been taken as $\varepsilon^{ME}=11.6\varepsilon_{0}$ and $20.5\varepsilon_{0}$ for the pyroelectric ferromagnet BaMnF$_4$, which correspond to the values measured along its $a$ and $b$ crystallographic axes, respectively. Dotted curves (a) and (b) are set as reference for $g=0$. Comparative results are shown for Rouard's method (RM) and the modified Nakayama (N) expression (Eq. \ref{Nakay}), indicating the change in the reflectivity spectra under the ME effect and different values for the dielectric constant $\varepsilon^{ME}$. The reflectance response increases from $0.4$ ($g=0.0$) to $0.63$ ($g=0.6878$) for electronic densities lower than $\sim 100\times 10^{13}$ cm$^{-2}$, while it augments monotonically to $1.0$ for electronic concentrations greater than $\sim 200\times 10^{13}$ cm$^{-2}$ regardless of the value of $g$, in the framework of the RM approach. One of the discrepancies with the Nakayama results is due to the difference between the 2D intrinsic plasma frequency $\Omega_S$ and those associated with the plasma frequency in the \emph{bulk} system $\omega_{P}$. Variation in the electronic carrier density in the former case has been simulated by inserting the thickness film dependence $\ell$ on $\omega_{P}$, providing good agreement for $\ell\sim 10$ nm ($\nu\sim 147.42$ cm$^{-2}$) as proven in Fig.(\ref{comp}).    
Minima of reflectivity obtained from Eq. (\ref{cuasiwv}), are located at $\lambda_{c}=2\pi\left(\varepsilon_{1}+\varepsilon_{2}\right)c/\Omega_{S}$,
or $\lambda_{c}^{-1}\propto\nu$, indicating that the critical
wavelength for \emph{bare} plasmon excitations is
larger as the electronic concentration decreases.  AFMR mode lies in the range THz range, with $\omega_{m}\sim
0.54$
THz, while the transverse phonon frequency is
taken as
7.53 THz for the BaMnF$_4$ compound\cite{Barnas}.
Metallic
behavior predominates for concentrations higher
than
$10^{16}$ cm$^{-2}$ and smaller than $10^{14}$ cm$^{-2}$
and selected wavelengths between 0.5 mm
and 0.6 mm. Resonant plasmon modes (i.e., collective electronic excitations under ME interaction) are
important for
carrier densities around $10^{15}$ cm$^{-2}$,
where
radiative absorption or antireflective phenomena
become strong and the reflectance spectrum is therefore
significantly modified by diminishing the
percentage of absorbed radiation only when the external
frequency approaches the characteristic mode $\omega_m$, and $g\neq 0$. 
Figure (\ref{r2}) depicts the shifting of the
minimum of reflectance in the $\left(\nu,g\right)$ plane for the Nakayama approach.
The ME effect becomes relevant by decreasing the
\emph{critical}
carrier density $\nu_c$ as $g$ increases, and it
remains essentially unmodified for those frequencies  away from
the AFMR characteristic mode as indicated in line (d). Dotted vertical line is tagged at $g=0.6878$ mm$^{-2}$ as a eye guide for identifying the critical density change as the incident wavelength varies around $2\pi c/\omega_m$. Critical density $\nu_{c}$ shall be understood as the electron carrier concentration which maximizes antireflective effects for the composite metal/multiferroic system. Figure (\ref{bparallel0}) shows the reflectance response under applied magnetic field with magnitude $1.5$ T for different directions on the $XY$ plane. AFMR resonance at $2\pi c/\omega_{m}$ is not essentially affected by the orientation of the external field, but it becomes sensitive with the azimuthal angle for frequencies between the edge of the THz range and the microwave (SHF) band. Highly reflective effects are more intense for external magnetic fields which are applied in the opposite direction with respect to the weak ferromagnetic state $\mathbf{M}$, favoring the metallic behavior for long wavelengths and shielding the resulting ME interaction.
\emph{In-plane}
applied field $\mathbf{B}$ effects on the
reflectance as a
function of carrier density $\nu$ are illustrated
in Fig.
(\ref{bparallel}). $R\left(\mathbf{B}\right)$
tends to increase for $\mathbf{B}$ parallel to $+X$-axis and decreases for $\mathbf{B}$ along $-X$ axis. Curve
(b) for
null $\mathbf{B}$ overlaps the outcome of $R$ at
$B=1.5$ 
T, $\phi=\pi /2$ and $\phi=\pi/2$ (i.e., parallel to $Y$ axis), indicating no substantial
variation in the
optical reflectance for applied fields in the same direction of the plasmonic wavevector $q_{Y}$ for carrier densities smaller than $\sim 10^{13}$ cm$^{-2}$. Equation (\ref{Max}) has also been treated by implementing Finite Element Method (FEM) and standard boundary conditions for $\mathbf{D}$ and $\mathbf{B}=\mu_{0}\mathbf{H}+4\pi\left[\mathbf{\chi}\right]\mathbf{E}$ fields in order to calculate the reflectance response as a function of incident wavelengths. 
Comparative results on the calculated response of the reflectance are shown in figures (\ref{comp}) and (\ref{3cc}). Under Nakayama's formalism, the metallic medium is treated as a 2D system, while Rouard and FEM methods converge with the first one for a film thickness around $\ell\sim 10$ nm, which roughly corresponds to an electronic carrier density of 147.42 cm$^{-2}$ after calculating the correlation between two intrinsic plasma frequencies $\omega_{P}$ and $\Omega_S$. Iso-reflective lines for $\Delta R/R=R\left(B\right)/R\left(0\right)-1$\cite{JDE} close to $2\pi c/\omega_{m}$ and the externally applied magnetic field (in $Z$ direction) are shown in Figure (\ref{control}). Projected lines preserve symmetrical distribution under magnetic field inversion nearby $\lambda_m$ although strong fluctuations and a sign flip on $\Delta R/R$ are present for wavelengths slightly different from $\lambda_m$ and magnetic fields greater than $\sim 5$ T, indicating that interacting ME and plasmonic activity might increase the reflectance outcome from systems with low electronic density and without applied field.     
\begin{figure}
\centering
\includegraphics[width=9cm,
scale=1]{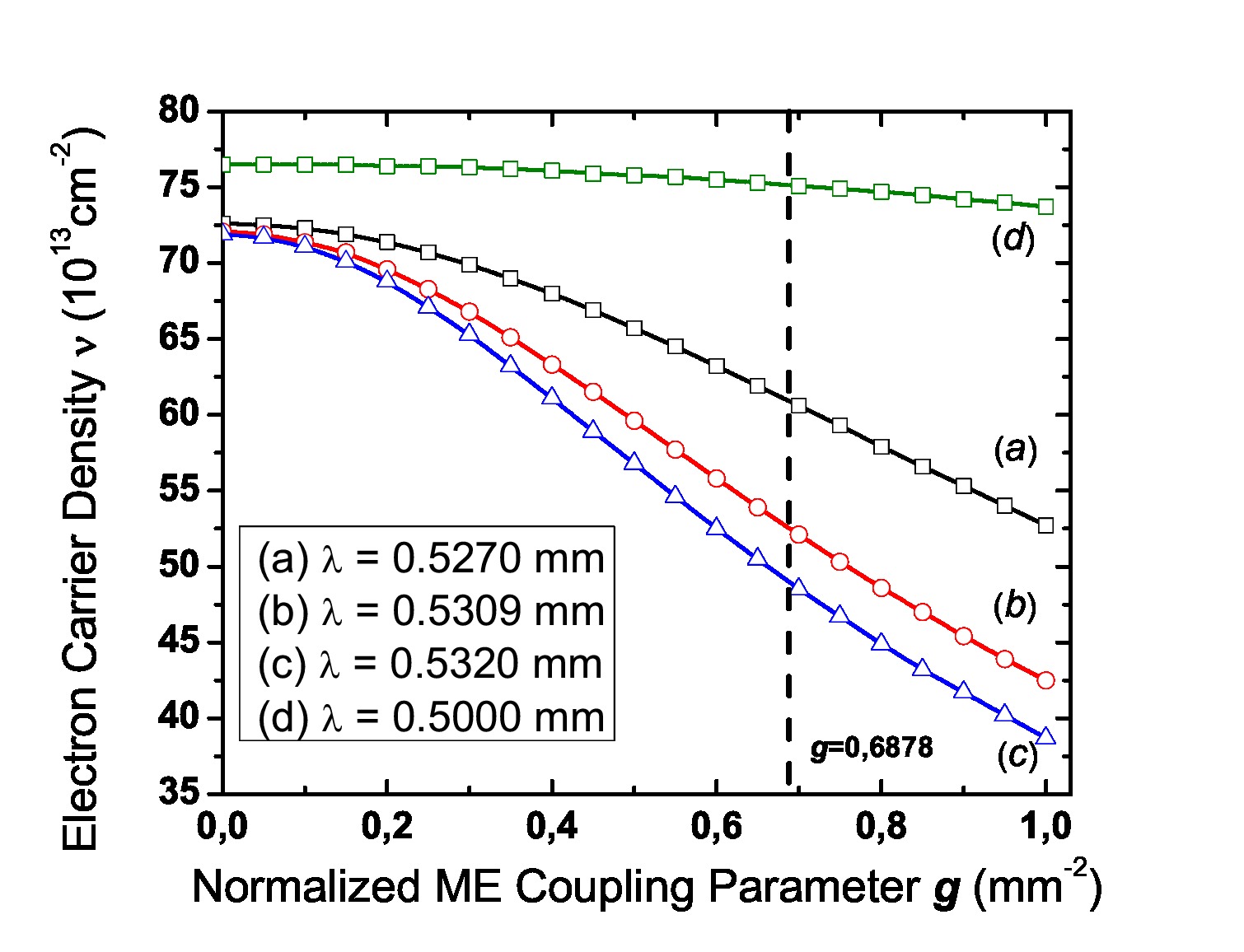}\label{r2}
\caption{Critical carrier density $\nu_{c}$ as a
function
of the ME coupling parameter $g$ for different
wavelengths. $\nu_c$ is strongly depending on $g$
only for
external frequencies near to AFMR mode
$\omega_m$.}\label{r2}
\end{figure}
\section{CONCLUDING REMARKS}
We have developed a model for studying the
magnetoelectric interactions on 2D plasmonic modes in the THz
range for a metal/multiferroic composite device. The multiferroic
medium exhibits weak ferromagnetism and the metallic
behavior enters into the formalism in the framework of the classical Drude-Lorentz model. Relative reflectance response for normal incidence is numerically calculated 
for a particular ME coupling strength $g$ and
wavelengths near to the optical antiferromagnetic resonance frequency
$\omega_{m}$ by using three different approaches: Nakayama's formalism, Rouard's method and Finite Elements (FEM). Characteristic soft phonon and AFMR frequencies were taken for the pyroelectric ferromagnet BaMnF$_4$, showing that a particular condition for reflectivity might
be adjustable by varying the intensity of the applied field, its orientability, film thickness or incident frequency of radiation, mainly in a range $\lambda>\lambda_{m}$. Spectra of reflectance demonstrate that the magnetoelectric interaction predominates for metallic film thicknesses smaller than 25 nm in the THz regime, while for thicker films (50-100 nm) the optical outcomes are not significantly affected by this interaction; instead, total reflectance from the film is observed along a wide range of frequencies up to the cut-off bulk value $\omega_{P}\sim 2.1$ PHz, in which reflectivity decays abruptly to zero and exhibits oscillatory behavior for greater frequencies. The chosen value of $\omega_P$ is into the typical order of magnitude for good conductors like gold, silver or copper, despite that the calculations and comparison with the strictly 2D system were made just for the first one. There is not a clear signature of the plasmonic cut-off for intermediate film thicknesses (25-50 nm) and the reflectivity curve does not breach abruptly as for wider ones; rather, it reaches its maximum value in a broad interval of $10^{13}-2\times 10^{16}$ Hz, suggesting a variation of the effective dielectric response associated with the metal under ME interaction.     
Further analysis shall be proposed for other metals or semiconducting materials, since optical control experiments on the THz range have recently been
achieved on GaAs  wafers via stimulated
photocarriers generated by interband light absorption.
The resulting reflectivity spectrum is tuned from antireflective ($R<3\%$) to high reflective
($R>85\%$) limits under controlled power illumination\cite{Fekete},\cite{Coutaz}. Although all numerical simulations were conducted
for $\varepsilon^{ME}=K\varepsilon_{0}$, ($K$ being taken as $11.6$ and $20.5$ in the range of interest), simultaneous electric field control
$\mathbf{E}_0$ on optical properties for the
composite device might also be achieved under the dielectric
function dependence for a multiferroic material
$\varepsilon_{2}\left[\mathbf{P}\left(\mathbf{E}_{0}\right)\right]$,
the polarization $\mathbf{P}\left(\mathbf{E}_{0}\right)$
and temperature, issue that shall be addressed in
future investigations.
\newline
\newline
\begin{acknowledgments}
H.V. wants to thank computing accessibility at
POM group.
C. V.-H. acknowledges financial support provided
by DIMA,
\emph{Direcci\'on de Investigaci\'on Sede
Manizales},
Universidad Nacional de Colombia. H.V. declares
no
competing financial interest.
\end{acknowledgments}

\end{document}